\documentclass[12pt,epsf]{article}

\newlength{\pubnumber} 

\catcode`\@=11
\@addtoreset{equation}{section}

\def\section{\@startsection{section}{1}{\z@}{3.5ex plus 1ex minus .2ex}
 {2.3ex plus .2ex}{\large\bf}}
\def\subsection{\@startsection{subsection}{2}{\z@}{2.3ex plus .2ex}
 {2.3ex plus .2ex}{\bf}}

\usepackage{graphicx}
\usepackage{dcolumn}
\usepackage{bm}
\usepackage{amssymb,amsmath}

\def\beq{\begin{equation}}
\def\eeq{\end{equation}}
\def\beqn{\begin{eqnarray}}
\def\eeqn{\end{eqnarray}}

 \font\cmss=cmss10 \font\cmsss=cmss10 at 7pt

\def\IZ{\relax\ifmmode\mathchoice
 {\hbox{\cmss Z\kern-.4em Z}}{\hbox{\cmss Z\kern-.4em Z}}
 {\lower.9pt\hbox{\cmsss Z\kern-.4em Z}}
 {\lower1.2pt\hbox{\cmsss Z\kern-.4em Z}}\else{\cmss Z\kern-.4em Z}\fi}

\def\Io{\relax\ifmmode\mathchoice
 {\hbox{\cmss 1\kern-.4em 1}}{\hbox{\cmss 1\kern-.4em 1}}
 {\lower.9pt\hbox{\cmsss 1\kern-.4em 1}}
{\lower1.2pt\hbox{\cmsss 1\kern-.4em 1}}\else{\cmss 1\kern-.4em 1}\fi}

\begin{document}
\begin{titlepage}
\setcounter{page}{1}
\rightline{BU-HEPP-08-15}
\rightline{CASPER-08-04}
\rightline{\tt }

\vspace{.06in}

\begin{center}
{\Large \bf Casimir Energy and Brane Stability}\vspace{.12in}

{\large
        R. Obousy\footnote{Richard\_K\_Obousy@baylor.edu}
        and G. Cleaver\footnote{Gerald\_Cleaver@baylor.edu}}
\\
\vspace{.12in}
{\it        Center for Astrophysics, Space Physics \& Engineering Research\\
            Department of Physics, Baylor University,
            Waco, TX 76798-7316\\}
\vspace{.06in}
\end{center}

\begin{abstract}

We investigate the role of Casimir energy as a mechanism for brane stability in five-dimensional models with the fifth dimension compactified on an $S^1/\IZ_2$ orbifold, which includes the Randall-Sundrum two brane model (RS1). 
We employ a $\zeta$-function regularization technique utilizing the Schwinger proper time method and the Jacobi theta function identity to calculate the one-loop effective potential. We show that the combination of the Casimir energies of a scalar Higgs field, the three generations of Standard Model fermions and one additional massive non-SM scalar in the bulk produce a non-trivial minimum of the potential. In particular, we consider a scalar field with a coupling in the bulk to a Lorentz violating vector particle localized to the compactified dimension. Such a scalar may provide a natural means of the fine-tuning needed for stabilization of the brane spearation. Lastly, we briefly review the possibility that Casimir energy plays a role in generating the currently observed epoch of cosmological inflation by examining a simple five-dimensional anisotropic metric.

\end{abstract}
\end{titlepage}
\setcounter{footnote}{0}


\section{Introduction}

The proposal that there exists extra spatial dimensions in which gravity and possibly other fields can propagate has been the subject of intense study of recent \cite{aadd,add,s,rs1,rs2}. Especially popular are the five-dimensional models with $S_1$ or $S_1/\IZ_2$ compactification. Of particular interest is the Randall-Sundrum model which proposes a novel geometrical solution to the heirachy problem (RS1)\cite{rs1}. In the RS1 setup the Standard Model (SM) fields, except for possibly the Higgs field H, are not simply confined to one of two 3-branes which lie at the endpoints (i.e., fixed points) of an $S^1/\IZ_2$ orbifold, but are extended along the orbifold. Nevertheless, one of the branes physically corresponds to `our' universe and is sometimes referred to as the IR or `visible' brane. The second brane is the UV or `hidden' brane. 
The line element in RS1 is described by the metric
\begin{equation}
ds^2=e^{-2kr_c|\varphi|}\eta_{\mu\nu}dx^\mu dx^\nu - r_c^2 d\varphi^2,
\label{eq1}
\end{equation}
where the points $(x^\mu,\varphi)$ and $(x^\mu,-\varphi)$ are identified with each other, $x^\mu$ are the standard four dimensional coordinates and $|\varphi|\leq\pi$. The exponential factor is referred to as the warp factor and is an appealing feature in the RS1 model, as it can generate a TeV mass scale from the Planck scale in the higher dimensional theory, while retaining a bulk width that is only a couple of orders of magnitude above the Planck scale. Consider for example the action of a Higgs field, 
\begin{equation}
S_{vis}=\int d^4x\sqrt{-g_{vis}}\left(g^{\mu\nu}_{vis}D_\mu H^\dagger D_\nu H-\lambda(|H|^2-v_0^2)^2\right),
\label{eq1a}
\end{equation}
where the $\lambda$ is a lagrange multiplier which ensures $|H|^2=v_0^2$, fixing the mass parameter of the Higgs. If we now substitute eq.\ (\ref{eq1}) into this action we find
\begin{equation}
S_{vis}=\int d^4x\sqrt{-g_{hid}}e^{-4kr_c\pi}\left(g^{\mu\nu}_{hid}e^{2kr_c\pi}D_\mu H^\dagger D_\nu H-\lambda(|H|^2-v_0^2)^2\right),
\label{eq1b}
\end{equation}
If we now redefine $H\rightarrow e^{kr_c\pi}H$ we see, 
\begin{equation}
S_{eff}=\int d^4x\sqrt{-g_{hid}}\left(g^{\mu\nu}_{hid}D_\mu H^\dagger D_\nu H-\lambda(|H|^2-e^{2kr_c\pi}v_0^2)^2\right).
\label{eq1c}
\end{equation}
The remarkable feature of this model is that a field with mass $m_0$ on the $\varphi=0$ brane will have a reduced physical mass of $m\approx \rm{e}^{-2\pi k r_c} m_0$ on the $\varphi=\pi$ brane. Typically $2\pi kr_c\approx 12$. In this model the branes themselves remain static and flat.

Naively one might expect an extra dimension to either contract to the Planck length, or inflate to macroscopic scales and so the question of stabilization becomes important. Negative energy is a vital component generic to realistic stabilization schemes \cite{adnv}. Negative-tension orientifold planes are one possible source of this negative energy in M-theory, another is the Casimir energy of fields in the bulk or in compact space.
In 1948 H. Casimir published a paper where he explained the van der Waals interaction in terms of the zero-point energy of a quantized field \cite{cas}. In its most elementary form the Casimir effect is the interaction of a pair of neutral parallel plates. The presence of the plates modifies the quantum vacuum and this modifcation causes the plates to be pulled toward each other.  

One attractive feature of the Casimir energy in stabilization schemes is that it is an inherent property of the quantum vacuum, and does not need to be added `by hand'. Additionally, the Casimir effect can easily be extended to regions of non-trivial topology \cite{aw,ke} adding to its theoretical attractiveness. For example, on $S^1$, a circular manifold, one can associate $0$ and $\pi$ with the location of the plates and the Casimir energy can be calculated. This becomes relevant when we consider models with additional spatial dimensions \cite{gl}.

Since the pioneering work of Appelquist and Chodos \cite{ac,ac1}, it has been known that the Casmir effect due to quantum gravitational fluctuations can generate a minimum of the vacuum potential in Kaluza-Klein (KK) models. This minimum prevents the extra dimension from continuing to either shrink or expand. Extensions of this work include demonstrating that massless and/or massive matter fields could stabilize the fifth dimension against collapse \cite{rr}. KK setups in which the extra dimension is an $S^1$ or an $S^1/\IZ_2$ topology is not the only extra dimensional scenario, and the utility of the Casimir energy as a stabilization mechanism has proved to be a rich field of research. For example toroidal topologies, $T\sim S^1 \otimes S^1$, have been examined \cite{I,bcp}, as have more sophisticated surfaces. More exotic spacetimes, such as Anti-de Sitter(AdS) space and brane world scenarios, have also been investigated \cite{wg1,ir1,fh,gpt,enoo,gpt1,nos,m,fmt,hkp,kt,nos1,ns,ss,Bai2008}.

Classical stabilization forces have also been investigated, for example Gell-Mann and Zweibach \cite{gmz} examined the stabilization effect due to a scalar field along an extra dimension. In this model the compactification mechanism uses a scalar sector in the form of a nonlinear sigma model with the action
\begin{equation}
S=  \frac{1}{2} \int d^{D+E}x\sqrt{g}\left(-\frac{R}{2}+\frac{g^{\mu\nu}}{\lambda^2}h_{ij}(\phi)\partial_\mu\phi^i\partial_\nu\phi^j\right),
\label{eq1d}
\end{equation}
where $(D+E)$ is the total number of spacetime dimensions. The scalar fields $\phi^i(x),i=1,2,3,...E$ are thought of as coordinates on an E-dimensional Riemannian manifold $\cal{M}_E$ with metric $h_{ij}(\phi)$. The number of scalar fields equals the number of dimensions that are to be made compact. 

The solutions to the equations of motion of this system demonstrate that the extra dimensions roll up to form a compact manifold with an internal volume determined by $\lambda$ which is a parameter of the classical solution. The nonlinear sigma model is not thought to be fundamental, but rather an effective field theory for composite scalars arising in a theory at some energy below the Planck mass. This work was expanded on by Goldberger and Wise \cite{gw}, who analyzed the classical stabilization forces in the context of brane worlds. However this proved not to be useful for the stabilization of two positive tension branes \cite{kop,bhlm}.

As well as possibly playing a role in the stabilization of higher dimensions, Casimir energy can also be investigated in the context of cosmology. One of the most profound results of modern cosmolgy is the evidence that 70\% of the energy density of the universe is in an exotic form of energy with negative pressure that is currently driving an era of accelerated cosmological expansion \cite{wv,d}. The results first published in \cite{p} have drawn huge interest as theoreticians attempt to explain the nature of this `dark' energy. For certain topologies it has been shown that positive Casimir energy leads to a non-singular de Sitter universe with accelerated expansion \cite{mt}, and possible links between Casimir energy and dark energy have been made in the literature \cite{gl,milt}. We therefore consider it to be worthwhile to briefly consider Casimir energy from a cosmological perspective. 

In Section 2  we begin by reviewing the derivation of the four-dimensional effective theory with a discrete KK tower from the scalar field. In Section 3 we derive in detail the Casimir energy for a periodic and anti-periodic massive scalar field using $\zeta$-function techniques. Although it is quite possible that this specific derivation has appeared in the literature, the authors have not come across it and so we show the calculation in its entirety. In Section 4 we review the derivation for the Casimir energy in the case of an exotic coupling of a massive scalar field to an antisymmetric Lorentz violating tensor. The  resulting expression for an enhanced Casimir energy was first derived by the authors in \cite{oc}, and will be investigated as a component in the stabilization scenarios we study. In Section 5 we explore which field combinations generate a minimum of the Casimir energy. Finally in Section 6 we discuss the role of the Casimir energy in the dynamics of cosmological evolution.

\section{Scalar Field in Randall-Sundrum Background}

In this section we review the equation of motion of a scalar field in the RS1 setup \cite{rs1}. In this scenario the heirachy between the electroweak scale and the Planck scale is generated by introducing a $5^{th}$ dimension compactified on an $S^1/Z^2$ orbifold with large curvature. 
Two 3-branes with opposite tension are located at the orbifold fixed points. The line element in RS1 is described by the metric eq.\ (\ref{eq1}). This review follows closely \cite{gw}.

Consider a free scalar field in the bulk
\begin{equation}
\mathcal{L}=  \frac{1}{2}G_{AB}\partial_A\Phi \partial_B\Phi-\frac{1}{2} m^2\Phi^2.
\label{eq2}
\end{equation}
Solving the equation of motion we obtain
\begin{equation}
e^{-2kR|\varphi|}\eta_{\mu\nu}\partial_\mu\Phi \partial_\nu\Phi+\frac{1}{R^2}\Phi\partial_\phi(e^{-4kR|\varphi|}\partial_\phi)-m^2e^{-4kR|\varphi|}\Phi^2=0.
\label{eq3}
\end{equation}
To separate out the extra dimensional contributions we first use separation of variables and express the field as
\begin{equation}
\Phi(x,\phi)=\sum_n\psi_n(x)\frac{y_n(\phi)}{\sqrt{R}},
\label{eq4}
\end{equation}
and find the equation for y to be
\begin{equation}
-\frac{1}{R^2}\frac{d}{d\phi}\left(e^{-4kR|\varphi|}\frac{dy_n}{d\phi}\right) +m^2e^{-4kR|\varphi|}y_n=m_n^2e^{-2kR|\varphi|}y_n.
\label{eq5}
\end{equation}
The bulk scalar manifests itself in four dimensions as a tower of scalars with respective masses $m_n$. To solve equation eq.\ (\ref{eq5}) it is useful to perform a change of variable, $z_n=m_n e^{kR|\varphi|}/k$ and $f_n=e^{-2kR|\varphi|}/y_n$. We can now write eq.\ (\ref{eq5}) as 
\begin{equation}
z_n^2\frac{d^2f_n}{dz_n^2}+z_n\frac{df_n}{dz_n}+\left[z_n^2-\left(4+\frac{m^2}{k^2} \right)\right]f_n=0.
\label{eq6}
\end{equation}
The solutions to this equation are Bessel functions:
\begin{equation}
y_n(\phi)=\frac{e^{2kR|\varphi|}} {N_n}\left[ J_\nu\left(\frac{M_ne^{kR|\varphi|}}{k}\right)+b_{n\nu}Y_\nu\left(\frac{M_ne^{kR|\varphi|}}{k}\right)\right].
\label{eq7}
\end{equation}
To satisfy the boundary conditions of \cite{gw} at $y=0$ and $y=\pi R$, the argument of the Bessel function has to satisfy
\begin{equation}
\frac{M_ne^{kR}}{k}\approx\pi(N+\frac{1}{4}),\ \ \ \ N\geq1.
\label{eq8}
\end{equation}
and we have a four-dimensional effective theory with a discrete KK spectrum for the scalar field with exponentially suppressed masses.

\section{Higher Dimensional Casimir Energy Calculations}

The Casimir energy generated from the quantum fluctuations in the large dimensions are insignificant when they are compared to the contributions arising from the compact dimensions because the energy is inversely proportional to volume of the space. Therefore, our first Casimir energy calculation focuses on the Casimir energy for a field with boundary conditions on the $S^1$ compactification, then modified for $S^1/\IZ_2$. We use $\zeta$-function techniques inspired by those discussed in the literature \cite{e1,e2,e3,kk1,e4,ekz,e5,i}.

For a massive field we can express the modes of the vacuum (using natural units) in RS1 as \cite{ftz}
\begin{equation}
E_n=\sqrt{\vec{k}^2+\left(\frac{\pi n}{r_c}\right)^2+M_n^2},
\label{eq9}
\end{equation}
with $M_n$ set by eq.\ (\ref{eq8}), and $r_c$ the radius of the compact extra dimension. The Casimir energy is given by
\begin{equation}
V^+=\frac{1}{2}{\sum_{n=-\infty}^{\infty}}' \int \frac{d^4k}{(2\pi)^4}{\rm log}(\vec{k}^2+\left(\frac{n\pi}{r_c}\right)^2+M_n^2),
\label{eq10}
\end{equation}
where the prime on the summation indicates that the $n=0$ term is excluded. For purposes of regularization we will write this as
\begin{equation}
V^+=\frac{1}{2}{\sum_{n=-\infty}^{\infty}}'\int\frac{d^4k}{(2\pi)^4}\int_0^\infty\frac{ds}{s}e^{-(\vec{k}^2+\left(\frac{n\pi}{r_c}\right)^2+M_n^2)s}.
\label{eq11}
\end{equation}
We first perform the Gaussian integration (the $k$-integral)
\begin{equation}
\int_0^\infty d^4ke^{-\vec{k}^2s}=\frac{\pi^2}{s^2},
\label{eq12}
\end{equation}
and are left with the remaining calculation;
\begin{equation}
V^+=\frac{1}{2}\frac{\pi^2}{(2\pi)^4} {\sum_{n=-\infty}^{\infty}}'\int_0^\infty\frac{1}{s^3}e^{-\left((\frac{n\pi}{r_c})^2+M_n^2\right)s}.
\label{eq13}
\end{equation}
To help us solve this equation we will use the Poisson Resummation formula (Jacobi's theta function identity),
\begin{equation}
{\sum_{n=-\infty}^\infty}' e^{-(n+z)^2t}=\sqrt{\frac{\pi}{t}}\sum_{n=1}^\infty e^{-\pi^2n^2/t}{\rm cos}(2\pi nz),
\label{eq14}
\end{equation}
to rewrite the summation of eq.\ (\ref{eq13}). Setting $z=0$, we obtain
\begin{equation}
\sum_{n=-\infty}^\infty e^{-(\frac{n\pi}{r_c})^2}=\sqrt{\frac{1}{\pi s}}\sum_{n=1}^\infty e^{r_c^2n^2/s}.
\label{eq15}
\end{equation}
Inserting this back into eq.\ (\ref{eq13}), we find
\begin{equation}
V^+=\frac{1}{2}\frac{\pi^2}{(2\pi)^4} \sum_{n=1}^\infty \sqrt{\frac{1}{\pi s}}e^{-(r_c^2n^2/s+M_n^2s)},
\label{eq16}
\end{equation} 
and so our expression for the Casimir energy density now becomes
\begin{equation}
V^+=\frac{1}{2}\frac{\pi^2}{(2\pi)^4} \sqrt{\frac{1}{\pi}}\sum_{n=1}^\infty \int_0^\infty ds \frac{1}{s^{7/2}}e^{-(M_nr_cn(\frac{M_ns}{r_cn}+\frac{r_cn}{M_ns}))}.
\label{eq17}
\end{equation} 
If we now set $x=\frac{M_ns}{r_cn}$, we can write eq.\ (\ref{eq17}) as
\begin{equation}
V^+=\frac{1}{2}\frac{\pi^2}{(2\pi)^4} r_c^{-5/2}M_n^{5/2}\sum_{n=1}^\infty \frac{1}{n^{5/2}}\int_0^\infty dx x^{-7/2}e^{-M_nr_cn(x+\frac{1}{x})}.
\label{eq18}
\end{equation} 
The integral is easily solved using the following expression for the Modified Bessel function of the Second Kind:
\begin{equation}
K_\nu(z)=\frac{1}{2}\int_0^\infty dx x^{\nu-1}e^{-z/2(x+\frac{1}{x})}.
\label{eq19}
\end{equation} 
Using eq.\ (\ref{eq19}) in eq.\ (\ref{eq18}) and recognizing the infinite sum as the Riemann zeta function we obtain our final expression for the Casimir energy density of a periodic massive scalar field in the five dimensional setup.
\begin{equation}
V^+=-\frac{\zeta(5/2)}{32\pi^2}\frac{M_n^{5/2}}{r_c^{5/2}}\sum_{n=1}^\infty K_{5/2}(2M_nr_cn).
\label{eq20}
\end{equation} 
It is straightforward to extend this expression to include antiperiodic fields. Recalling eq.\ (\ref{eq14}) we see that for antiperiodic fields we can make the substition $n \rightarrow n+1/2$ which ensures the summation is over integer multiples of 1/2. This implies our z term in the Poisson Resummation fomula is now non-zero $(z=1/2)$, so we simply have to include the cosine term in our final Casimir energy expression. Thus, the Casimir energy for anti-periodic scalar fields in our five dimensional setup becomes
\begin{equation}
V^-=-\frac{\zeta(5/2)}{32\pi^2}\frac{M_n^{5/2}}{r_c^{5/2}}\sum_{n=1}^\infty K_{5/2}(2M_nr_cn){\rm cos}(n \pi).
\label{eq21}
\end{equation}
We now wish to find an expression of the Casimir energy due to a massless scalar which will also be used as a component in the stabilization investigation. The necessary calculation is
\begin{equation}
V^+_{massless}=\frac{1}{2}{\sum_{n=-\infty}^{\infty}}'\int\frac{d^4k}{(2\pi)^n}log(k^2+\left(\frac{n\pi}{r_c}\right)^2)
\label{eq22}
\end{equation}
This calculation is well known in the literature and so we simply quote the result
\begin{equation}
V^+_{massless}=-\frac{3\zeta(5)}{64\pi^2}\frac{1}{r_c^4},
\label{eq23}
\end{equation}
which is the limit of eq.\ (\ref{eq20}) as $M_{n}\rightarrow 0$.

The $\IZ_2$ constraint requires that we identify points on a circle related by the reflection $y=-y$.
Neglecting any brane contributions, the $S^1/Z^2$ orbifolding simply forces us to ignore all modes odd under this reflection, which means we discard half of the modes in the summation  eq.\ (\ref{eq10}) \cite{pp}. Our final expressions for the Casimir energy are thus simply multiplied by a factor of $\frac{1}{2}$.

From the expression for the Casimir contribution for a periodic massive scalar field it is straightforward to enumerate the Casimir contributions of all other massive and massless fields by using knowledge of five-dimensional supersymmetry multiplets \cite{pp},
\begin{equation}
V^+_{fermion}(r)=-4V^{+}(r),
\label{eq23a}
\end{equation}
\begin{equation}
V^-_{fermion}(r)=\frac{15}{4}V^{+}(r),
\label{eq23b}
\end{equation}
\begin{equation}
V^+_{higgs}(r)=2V^{+}(r),
\label{eq23c}
\end{equation}
\begin{equation}
V^+_{vector}(r)=3V^{+}(r),
\label{eq23d}
\end{equation}
\begin{equation}
V^+_{gravitino}(r)=-8V^{+}(r),
\label{eq23e}
\end{equation}
where the positive sign on the potential indicates a periodic field and a negative sign indicates an antiperiodic field.
Before we investigate which field combinations give stable minima it will be useful to introduce one additional ingredient into the model which will be explored in the next section.

\section{Casimir Energy of Scalar Field Coupled to an Exotic Lorentz Violating Fields}

In this section we apply the Casimir energy contriubtions reviewed in Section Three to a scalar field coupled to a vector field with a VEV localized to the $5^{th}$ dimension. Although this scenario clearly violates Lorentz invariance, it has been well explored in the literature \cite{k,kp}, and tests of Lorentz invariance violations are receiving a lot of attention for a possible role in cosmology \cite{mnpss,fhl,kdm,acnp,b,fnpa}. It was demonstrated by Kostelecky that spontaneous Lorentz breaking may occur in the context of some string theories \cite{ks}. In the SM, spontaneous symmetry breakdown occurs when symmetries of the Lagrangian are not obeyed by the ground state of the theory. This can occur when the perturbative vacuum is unstable. The same ideas apply in covariant string theory which, unlike the SM, typically involve interactions that could destabilize the vacuum and generate nonzero expectation values for Lorentz tensors (including vectors)\cite{ck}.

A simple mechanism to implement local Lorentz violation is to postulate the existence of a tensor field with a non-zero expectation value which couples to matter fields. The most elementary realization of this is to consider a single spacelike vector field with a fixed norm. This field selects a `preferred' frame at each point in spacetime and any fields that couple to it will experience a local violation of Lorentz invariance.

Recently the authors calculated the Casimir energy for the case of a scalar field coupled to a field localized only to the $5^{th}$ dimension \cite{oc}. One novel feature of the setup was the demonstration that it allowed different spacings in the KK towers \cite{ct}, and consequentially the generation of an enhanced Casimir energy. By including this feature in our stabilization scenario, we increase the freedom to generate the minimum of potential which will be controlled by a single parameter that encodes the ratio of the 5th dimensional field VEV to the mass parameter.

The model is set in a five dimensional spacetime, with a spacelike five-vector $u^a=(0,0,0,0,v)$ field, which ensures four dimensional Lorentz invariance is preserved. The fifth dimension is compactified on a circle $S^1$ (or $S^1/\IZ_2$). 
If we first define the antisymmetric tensor $\xi^{ab}$  in terms of $u^a$
\begin{equation}
\xi_{ab}=(\nabla_a u_b -\nabla_b u_a)
\label{eq25},
\end{equation}
we can form the following action \cite{ct}: 
\begin{equation}
S=M_*\int d^5x \sqrt{g} \left[-\frac{1}{4}\xi_{ab}\xi^{ab}-\lambda(u_au^a-v^2)+\sum_{i=1} \mathcal{L}_i\right].
\label{eq26}
\end{equation}
Here the indices $a, b$ run from 0 to 4. $\lambda$ is a Lagrange multiplier which ensures $u^au_a=v^2$, and we take $v^2>0$. The $\mathcal{L}_i$ can represent various interaction terms. For this letter we will only investigate interactions with a scalar field. This form of the Lagrangian ensures the theory remains stable and propagates one massless scalar and one massless pseudoscalar \cite{dgb}. Of interest is the KK tower generated by the field in the compact dimension. 

Following \cite{ct}, a real scalar field $\phi$ is coupled to $u^a$, with the VEV of $u^a$ in the compactified extra dimension. The Lagrangian is
\begin{equation}
\mathcal{L}_{\phi}=  \frac{1}{2} (\partial\phi)^2 -\frac{1}{2} m^2\phi^2-\frac{1}{2\mu^2_\phi}u^a u^b\partial_a\phi\partial_b\phi.
\label{eq27}
\end{equation}
The mass scale $\mu_\phi$ is added for dimensional consistency. The background solution has the form $u^a=(0,0,0,0,v)$ which ensures four dimensional Lorentz invariance is preserved. With the addition of this field, the mass spectrum of the KK tower is modified by a non-zero $\alpha_\phi$: 
\begin{equation}
m^2_{KK}=k^2+(1+\alpha_\phi^2) \left( \frac{n\pi}{r_c} \right) ^2
\label{eq28}
\end{equation}
where $\alpha_\phi=\frac{v}{\mu_\phi}$ is the ratio of the 5th dimensional field VEV to the mass parameter.

With the expression for the KK modes of the scalar field found, we now turn to the calculation of the Casimir energy of the scalar field obeying periodic boundary conditions compactified on $S^1$ and interacting with the vector field $u^a$.   
\begin{equation}
V_{l.v}=\frac{1}{2}{\sum_{n=-\infty}^{\infty}}'\int\frac{d^4k}{(2\pi)^4}{\rm log}\left( k^2+(1+\alpha_\phi^2)\left(\frac{n\pi}{r_c} \right)^2,         \right), 
\label{eq29}
\end{equation}
where the prime on the summation indicates that the $n=0$ term is omitted. The final expression for the Casmir energy in an $S^1/\IZ_2$ orbifold for a scalar field with periodic boundary conditions coupled to a Lorentz violating vector field is
\begin{equation}
V_{l.v}=-\frac{3(1+\alpha^2_\phi)^2}{64\pi^2}\frac{1}{r_c^4}\zeta(5).
\label{eq30}
\end{equation}
This result indicates that the Casimir effect is enhanced by a factor of $(1+\alpha^2_\phi)^2$ in the presence of a Lorentz violating field. We shall see that this enhancement can play a role in generating a stable minima of the extra dimension.


\section{Investigating Higher Dimensional Stability}

We now explore the possibility of stabilization scenarios which involve the fields we have discussed. The basic ingredients will be the Casimir energy density of periodic massive scalar fields $V^{+}$ (e.g., the Higgs), an exotic periodic scalar field with a coupling to a Lorentz violating vector in the extra dimension $V^{+}_{lv}$, and massive periodic fermionic fields $\tilde{V}^{+}_i$. 
Because we are phenomenologically motivated, we choose the fermion field masses to be those of the three generation of the SM. Once we add the Casimir energy density contributions from the SM fields, we investigate which additional fields are necessary to generate a stable minimum of the potential. The masses of the SM fields can all be found in Appendix A.

\subsection{Standard Model Fields}

The first scenario we investigate involves populating the extra dimension with the SM fermionic fields
\begin{equation}
\tilde{V}^{+}_{ferm}\equiv\sum_{i=1} \tilde{V}^{+}_{i},
\label{eq31}
\end{equation} 
where the index i runs over all of the SM fermionic fields, apart from the left-handed antineutrino, and for which the masses are given in Appendix A. We also include the contribution from a bosonic Higgs-like field $V^{+}_{higgs}$. For computation of $V^{\rm tot}(r)$, we have normalized the masses of the SM particles in terms of the Z-boson mass. The top quark provides the majority of the contribution to the total mass of the fields. The potential is plotted in Figure 2 as a function of the radius of the fifth dimension. We investigate three possible Higgs masses; 115, 150 and 200 GeV. Here the lower limit is based on accelerator evidence (or lack there of), and the upper limit is based on theoretical predictions. 

Our expression for the total Casimir energy density is given by
\begin{equation}
V^{\rm tot}(r)=\tilde{V}^{+}_{ferm} + V^{+}_{higgs}, 
\label{eq32}
\end{equation} 
where our energy densities are calculated using eq.\ (\ref{eq20}) and eq.\ (\ref{eq23a}). For computation of $V^{\rm tot}(r)$, we have normalized the masses of the SM particles in terms of the Z-boson mass. The top quark provides the majority of the contribution to the total mass of the fields. The potential is plotted in Figure 1 as a function of the radius of the fifth dimension.


\begin{figure}[!ht]
\begin{center}
\includegraphics[width=500pt]{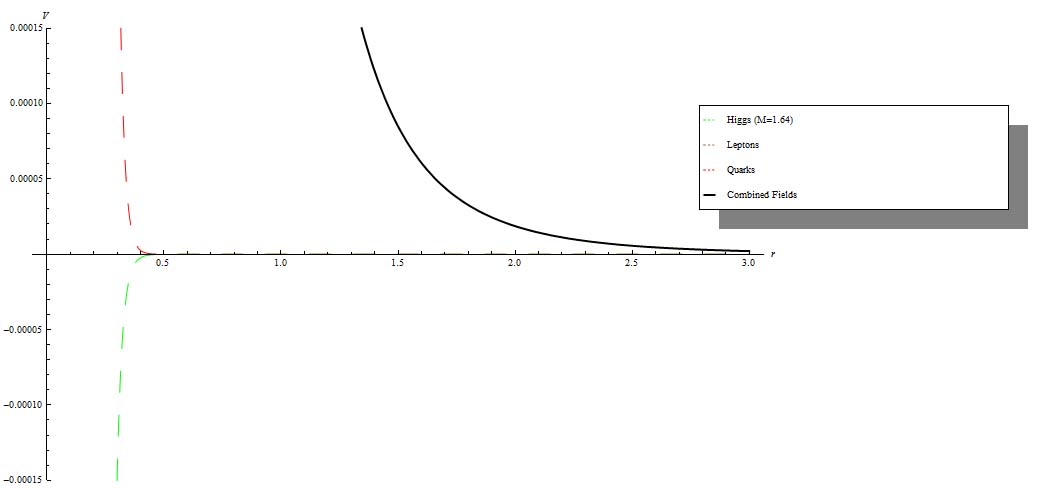}
\caption{\textit{The total contribution to the Casimir energy due to the standard model fermions and the Higgs field. The variation in Casimir energy density for the three values of the Higgs mass is shown, however the change is so minute that it cannot be discerned from the single solid black line, which also hides the contribution from the leptons also behind the black line. No stable minimum of the energy density is found with this field configuration.}}
\end{center}
\end{figure}

We find that that no stable minimum develops for this specific combination of fields and that the range of Higgs values has negligible bearing on the overall shape of the Casimir potential. We conclude that additional field contributions are necessary for the generation of a stable minimum.

\subsection{SM Fields, a Higgs Field and an Additional Exotic Massive Fermion}

Because these field contributions alone are not adequate to generate a minimum of the potential we add a contribution $\tilde{V}^{-}_E$ from some exotic antiperiodic massive fermionic field for which the mass is selected `by hand' to ensure a stable minimum.\footnote{This has some phenomenological motivations. For example, work by Mohapatra and others have motivated the possibility of a `light sterile bulk neutrino' as an explanation for solar and atmospheric neutrino oscillations \cite{cmy,cmy1,ddg,addm}.  The premise here is to postulate the existence of a gauge singlet neutrino in the bulk which can couple to leptons in the brane. This coupling leads to a suppression of the Dirac neutrino masses and is largely due to the large bulk volume that suppresses the effective Yukawa couplings of the KK modes of the bulk neutrino to the fields in the brane.}
Our expression for the total Casimir energy density is given by
\begin{equation}
V^{\rm tot}(r)=\tilde{V}^{+}_{ferm} + V^{+}_{higgs}+\tilde{V}^-_E, 
\label{eq34}
\end{equation} 
where our energy densities are calculated using eq.\ (\ref{eq20}) and eq.\ (\ref{eq23a}). 

\begin{figure}[!ht]
\begin{center}
\includegraphics[width=500pt]{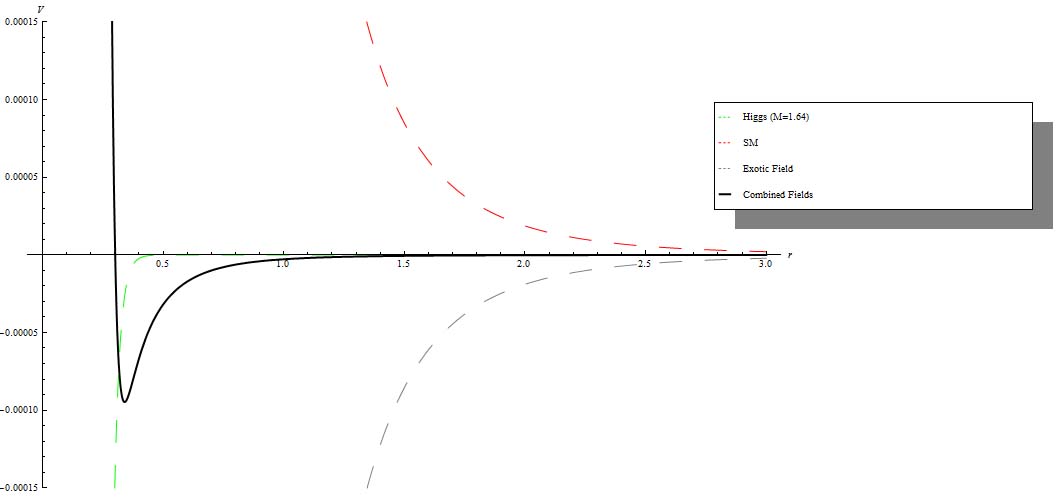}
\caption{\textit{With the addition of an antiperiodic massive fermionic field we see that a stable minimum of the potential develops.}}
\end{center}
\end{figure}

We find that with the inclusion of the SM fermions (including anti-particles except the left handed antineutrino) and the Higgs field, that a stable minimum develops when we include an additional contribution from a massive \textit{antiperiodic} fermionic field having a mass of $m=0.02$. The stable minimum in this scenario is negative and therefore corresponds to an AdS solution, however an additional positive contribution from the brane tension can easily be added to raise the overall potential above zero so that the minimum sits in a region of positive potential, thus generating a deSitter space.

Analysis of Figure 1 demonstrates that if the radius r is less than the critical value of $r=0.4$, the extra dimensions tends to grow, however as $r \rightarrow 0.4$, this growth is supressed and the extra dimensions is stabilized. Conversely if we start with a radius higher than the critical value, the extra dimension tends to shrink until the minimum is reached. Once the size of the $5^{th}$ dimension is stabilized the large dimensions experience increasingly more Casimir energy as they continue to expand, which is a salient feature of dark energy.

\subsection{Higgs Field and a Massless Scalar Field Coupled to Lorentz Violating Vector}

For our next study we consider the case of the Higgs field and a single massless scalar field with coupling to a Lorentz violating vector field of the type discussed in Section 4.  We explore the case of the massless scalar being both periodic and antiperiodic. With this choice of fields our expression for the Casimir energy in the compact fifth dimension becomes:
\begin{equation}
V^{\rm tot}(r)=V^{+}_{higgs}+(1+\alpha_\phi^2)^2V^{\pm}_{massless},
\label{eq34}
\end{equation} 
which we plot in Figure 3 for the case of a periodic field and Figure 4 for the antiperiodic fields. We also include the contributions for a range of coupling parameters $\alpha_\phi$.

\begin{figure}[!ht]
\begin{center}
\includegraphics[width=500pt]{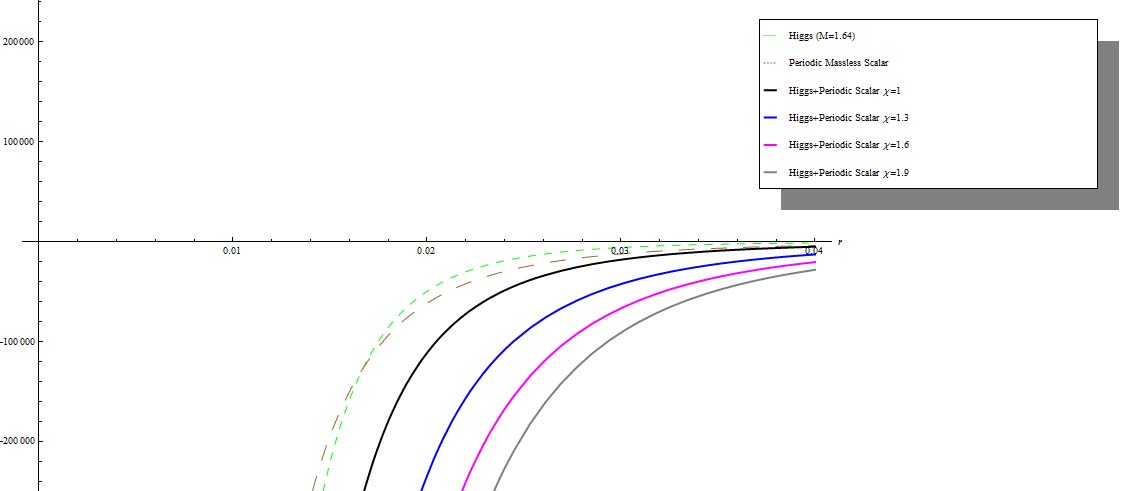}
\caption{\textit{The image illustrates a scenario with a Higgs field and a periodic massless scalar coupled to a Lorentz violating vector. It is clear that no stable minimum occurs for this choice of fields. The coupling parameter is encoded via $\chi=(1+\alpha_\phi^2)^2$.}}
\end{center}
\end{figure}

It is clear from Figure 3 that no stable minimum is obtained for the case of a Higgs field and a periodic massless scalar enhanced by $\chi$. This is because all the fields in this scenario contribute a negative Casimir energy and therefore there are no compensating positive contributions which would allow for the creation of the stable minimum. This vacuum is pathological and has no finite minimum at finite r. The energy density drops off aymptotically for all values of $\alpha_\phi^2$.

\begin{figure}[!ht]
\begin{center}
\includegraphics[width=500pt]{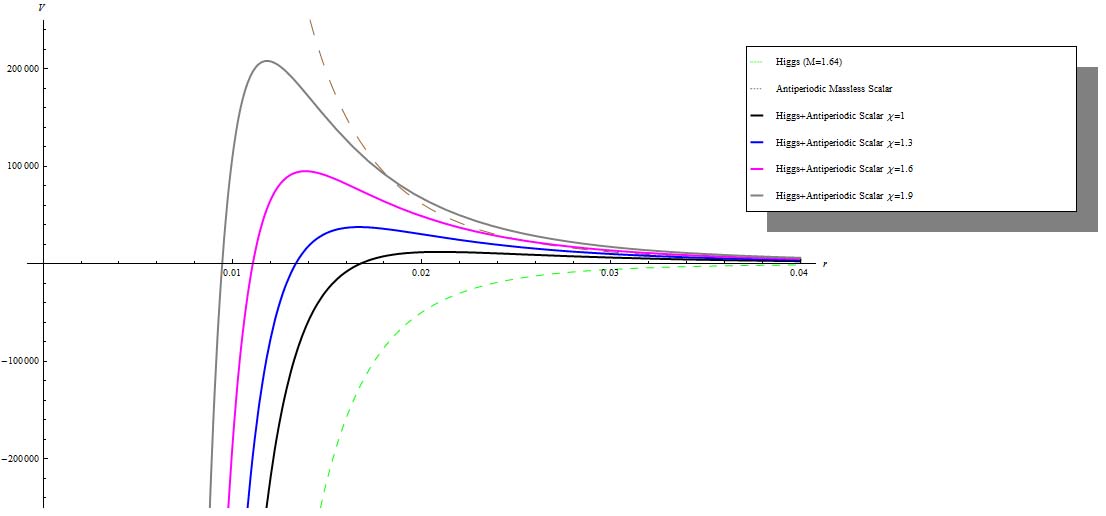}
\caption{\textit{This image illustrates a scenario with a Higgs field and an antiperiodic massless scalar coupled to a Lorentz violating vector. Again, no stable minimum occurs for this choice of fields. The coupling parameter is encoded via $\chi=(1+\alpha_\phi^2)^2$. }}
\end{center}
\end{figure}

However, for the periodic scalar shown in Figure 4, we find an unstable vacuum. If the radius begins at a distance less than the critical point, then from the perspective of an observer located in the bulk, a vacuum in the spacetime manifold would first be nucleated and then expand close to the speed of light. See for example discussions on false vacuum decay by Fabinger and Horava \cite{fh}.

For the universe to be deSitter in this scenario, the branes would have to start out separated by a distance above the critical value of around 0.015. The branes would steadily roll down the potential and the brane separation would grow larger as the Casimir energy density decreased. In this setup the possibility of vacuum tunnelling through the maximum exists, and this situation would represent a catastrophic fall into an ADS space. Such instabilites were studied in a KK scenario by Witten \cite{w}.

\subsection{Higgs Field, Standard Model Fermions and a Massless Scalar Field Coupled to Lorentz Violating Vector}

In this study we analyze the case of a Higgs field, the standard model fermions and a single massless (anti)periodic scalar field with coupling to a Lorentz violating vector field. With these fields our Casimir energy in the compact fifth dimension becomes:
\begin{equation}
V^{\rm tot}(r)=\tilde{V}^{+}_{ferm} + V^{+}_{higgs}+(1+\alpha_\phi^2)^2V^{\pm}_{massless}, 
\label{eq35}
\end{equation} 
which we plot in Figure 5 for periodic and in Figure 6 for antiperiodic massless scalar fields. 

\begin{figure}[!ht]
\begin{center}
\includegraphics[width=500pt]{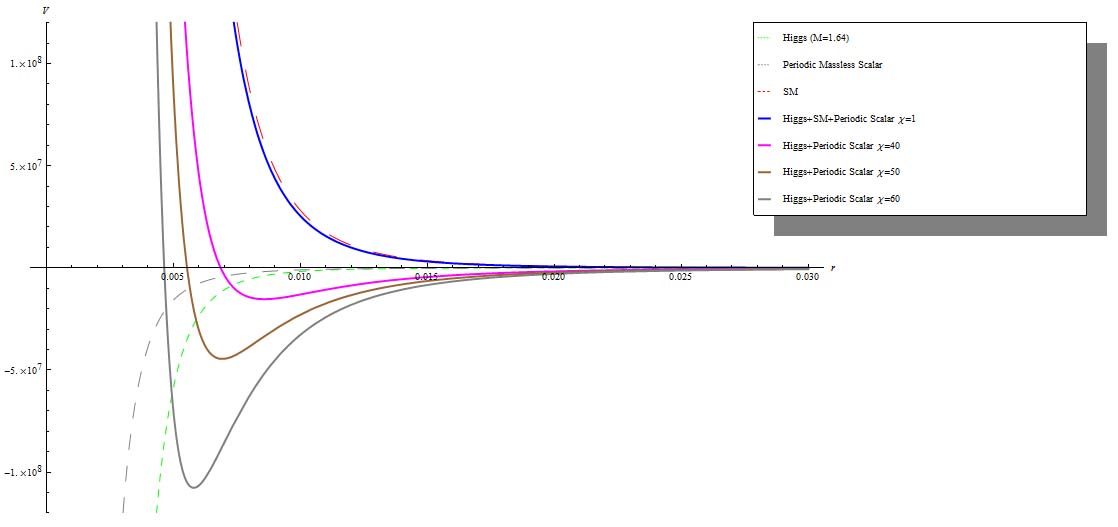}
\caption{\textit{In this image we show the contributions to the Casimir energy density from a Higgs field, the Standard Model fields and a periodic scalar field coupled to a Lorentz violating vector. As the parameter $\chi$ is increased the minimum of the potential is also decreased as is the radius of dimensional stabilization. Again here $\chi=(1+\alpha_\phi^2)^2$.}}
\end{center}
\end{figure}

\begin{figure}[!ht]
\begin{center}
\includegraphics[width=500pt]{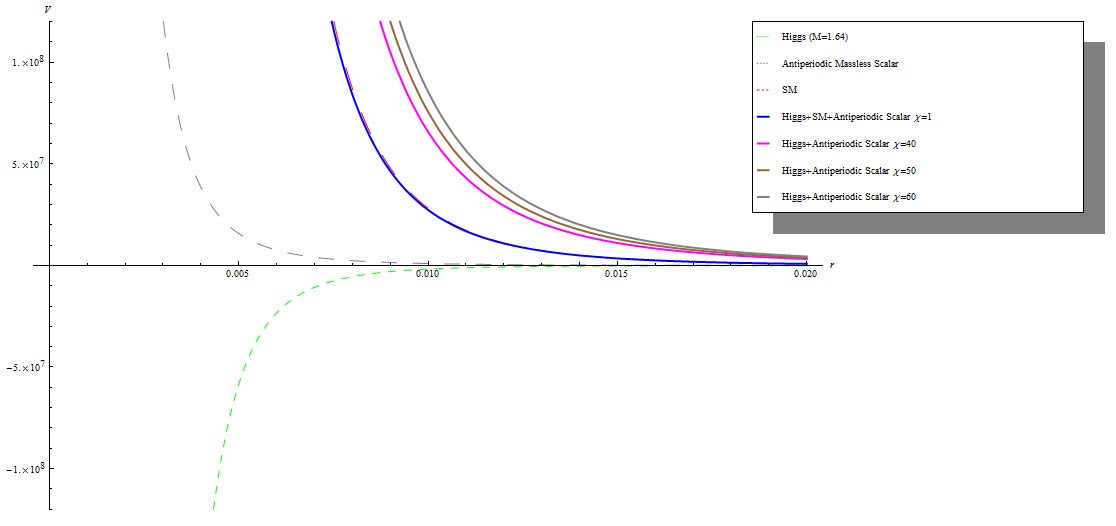}
\caption{\textit{Here we show the Casimir energy density with a Higgs field, the Standard Model fields and an antiperiodic scalar field coupled to a Lorentz violating vector. A stable minimum of the potential is not achieved for any value of $\chi$.}}
\end{center}
\end{figure}

We see that in the case of a periodic massless scalar field (Figure 5), as $\alpha_\phi^4$ is increased the Casimir energy is enhanced and consequentially the depth of the minimum increases while the stable minimum is located at progressively smaller radii. The nice feature of this field contribution is that the coupling parameter $\alpha_\phi$ is proportional to the VEV of the Lorentz-violating field. Thus, stabilization could correspond to minimization of a potential by this VEV. Hence, the apparant fine-tuning becomes a natural outcome. This is an advantage of having a massless scalar field that is coupled to a Lorentz-violating field. 

The case of an antiperiodic field is illustrated in Figure 6. We can see that even with a range of values of $\chi$, that no stable minimum is created.

\newpage

\subsection{DeSitter Minimum}

The minimums so far discussed have all been located at a negative Casimir potential indicating an AdS space. Because we know that our universe is expanding we now discuss a scenario which returns a positive minimum. For this we need at least two additional exotic fields and our expression for the Casimir energy in the compact fifth dimension becomes:

\begin{equation}
V^{\rm tot}(r)=\tilde{V}^{+}_{ferm} + V^{+}_{higgs}+\tilde{V}^{-}_E+V^+_E, 
\label{eq34}
\end{equation}
where the first exotic field, $\tilde{V}^{-}_E$ is an antiperiodic fermion and the second exotic field $V^+_E$ is simply a massive periodic scalar field. In this example the mass of $\tilde{V}^{+}$ is chosen to be 1.1 and the mass of $V^+_E$ is selected to be 1.8. Decreasing the mass of either of the exotic fields results in the minimum becoming deeper. It is clear from Figure 7 that a stable minimum of the Casimir energy density is created and that the minimum is located at $V(r)>0$ demonstrating a dS minima. The nice feature of this model is that mass of the mass of the exotic particles can be tuned to create a minimum that lies extremely close to the zero potential giving us a way recreate the accepted value of $\Lambda$.

\begin{figure}[!ht]
\begin{center}
\includegraphics[width=500pt]{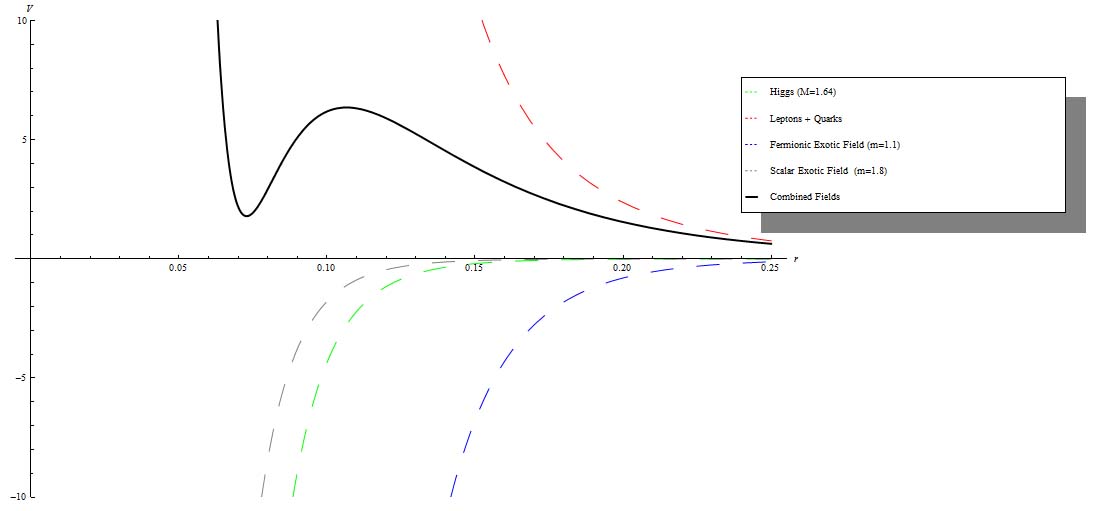}
\caption{\textit{ This field configuration consists of a Higgs field, the SM fields, an antiperiodic exotic massive fermionic field and a massive periodic scalar field. The minimum of the Casimir energy sits at a positive energy density indicating a deSitter spacetime.}}
\end{center}
\end{figure}

This field configuration allows the possibility of vacuum tunnelling out of the minimum ($r>0.11$) leading to an eternally inflating extra dimension.  

\section{Cosmological Dynamics}

We now briefly switch our discussion to examination of some the implications of generic higher dimensions in a cosmological context. We will focus on an d+n+1 dimensional anisotropic metric for simplicity. We consider a toy universe in which all the energy density content is due to Casimir energy contributions from the higher dimensions. We will see that in this setup, the Casimir energy density can, under certain conditions lead to an accelerated expansion scenario in the three large spatial dimensions. Following as in \cite{gl}, we start by considering a homogeneous and anisotropic metric 
\begin{equation}
ds^2=-dt^2+a(t)^2d\vec{x}^2+b^2(t)d\vec{y}^2
\label{eq32}
\end{equation} 
where a(t) and b(t) are the scale factors in the three large dimensions and the compact dimensions. We obtain the equations of motion  by varying the d+1 dimensional Einstein-Hilbert action, 
\begin{equation}
S=\int d^4xdy\sqrt{g}\left(\frac{M^{3}}{16\pi}R_{5}-\rho_{}\right)
\label{eq38}
\end{equation} 
with respect to the five-dimensional metric $g_{ab}$, from which we obtain the Einstein equations
\begin{equation}
3H_a^2+3nH_aH_b+\frac{n}{2}(n-1)H_b^2=8\pi G \rho_{}
\label{eq39}
\end{equation} 
\begin{equation}
\dot{H}_a+3H_a^2+nH_aH_b=\frac{8\pi G }{2+n}\left( \rho_{}+(n-1)p_a-np_b\right)
\label{eq40}
\end{equation} 
\begin{equation}
\dot{H}_b+nH_b^2+3H_aH_b=\frac{8\pi G }{2+n}\left( \rho_{}+2p_b-3p_a\right)
\label{eq41}
\end{equation} 

Let us first consider  eq.\ (\ref{eq35}) and analyse the simple scenario where the extra dimension has already found its minimum of potential implying $H_b=0$. Using the relation
\begin{equation}
\dot{H}_a=\frac{\ddot{a}}{a}+\left(\frac{\dot{a}}{a}\right)^2
\label{eq41b}
\end{equation} 
we find that eq.\ (\ref{eq35}) becomes
\begin{equation}
\frac{\ddot{a}}{a}+4H_a^2=\frac{8\pi G }{2+n}\left( \rho_{6D}+(n-1)p_a-np_b\right)
\label{eq42}
\end{equation}
and then using eq.\ (\ref{eq34}) we obtain
\begin{equation}
\frac{\ddot{a}}{a}=-\frac{8\pi G }{2+n}\left[ \left(\frac{5+4n}{3}\right)\rho_{6D}  +(n-1)p_a-np_b\right]
\label{eq43}
\end{equation}
For the case of n=2 we see that this equation simplifies to
\begin{equation}
\frac{\ddot{a}}{a}=-\frac{8\pi G}{4} \left[ \frac{13}{3}\rho_{6D} + p_a-2p_b\right]
\label{eq44}
\end{equation}
which implies that our current epoch of cosmological acceleration requires
\begin{equation}
\rho_{6D}\geq\frac{1}{13}\left(3p_a-6p_b\right)
\label{eq45}
\end{equation}

\section{Discussion}

We have investigated the possibility of moduli stability using the Casimir effect in RS1. We have calculated the one loop corrections arising from a massive scalar field with periodic boundary conditions in the compactified extra dimension by applying the Schwinger proper time technique and exploiting the Jacobi Theta function. We have populated the bulk with numerous fields in an attempt to uncover stabilization scenarios with an emphasis on phenomenologically motivated field content. Extending on previous work we have explored the implications of the existence of a five-dimensional vector field with a VEV in the compact dimension coupling to one of the scalar fields and noted its relevance as a tuning parameter. 

We have demonstrated that the Casimir energy of the SM fields in conjunction with the Higgs field cannot provide the necessary potential to stabilize the extra dimension. The fermionic nature of the SM fields contribute a \textit{positive} Casimir energy which is not balanced by the Higgs field when we include the three lepton generations (and their anti-particles, excepting the left-handed antineutrinos) and the six quarks and three color degrees of freedom(and their anti-particles). We have also investigated the possibility of stability for a range of Higgs masses based on experimental lower limits and theoretical upper limits and found the same result.

We have investigated the possibility of adding an exotic massive anti-periodic fermionic field to this field setup and discovered that a light field (m=0.02 in normalized units) is sufficient to generate a stable minimum of the potential. The minimum is located at a negative energy density which corresponds to an AdS spacetime. Reduction of the mass of the exotic field causes the minimum to become deeper. Our justification for the addition of this field is the possibility of the existence of a light sterile bulk neutrino.

Next we build on previous work by considering the effects on the Casimir energy of a scalar field coupled to a Lorentz violating vector field. This field, which is completely charactered by the parameter $\chi$, allows for fine tuning of the Casimir energy and the stabilization radius. Fine tuning is a generic feature of stabilization schemes and in this model the simple addition of a vector field in \textit{only} the fifth dimension creates additional freedom for the stabilization schemes. We discover that no stable minimum of the potential can be found with either a single periodic or antiperiodic massless scalar field coupled to a Lorentz violating vector in the case of a Higgs vacuum. However when the SM fields are included a periodic massless scalar field coupled to a Lorentz violating vector \textit{can} lead to a stable minimum, but this is dependent on the parameter $\chi$. Our motivations for studying the phenomenology of Lorentz violating fields stem from the recent surge of activity regarding the possibility Lorentz invariance violations and the potential role of Lorentz violating fields in cosmology.

We also outline a higher dimensional field configuration which creates a positive energy density minimum of the Casimir energy. We find that at least two additional exotic fields are required. Our example highlights a possible connection between dark energy, the heirachy problem and additional bulk fields. The capability of this model to explain so many apparantly unrelated phenomenon under the common framework of extra dimensional boundary conditions makes this model particularly appealing.

Their has also been recent interest in the literature relating dark energy to Casimir energy and for this reason we have briefly reviewed cosmological aspects of extra dimensions by considering an anisotropic cosmology. Using simple arguments we have found the relation between the Casimir energy density and the pressure in both large and compact dimensions necessary for accelerated cosmological expansion. 

\section{Acknowledgements}

Research funding leading to this manuscript was partially provided by Baylor URC grant 0301533BP.

\section{Appendix}

\begin{table}[ht]
\centering 
\begin{tabular}{c c c} 
\hline\hline 
Particle & Mass (GeV) & Normalized Mass \\ [0.5ex] 
\hline 
$e^-$     &$0.000511$        &$5.6\time10^{-6}$\\
$\mu^-$   &$0.106$           &$1.2\time10^{-3}$\\
$\tau^-$  &$1.78$            &$2.0\time10^{-2}$\\
$\nu_e$   &$1\times 10^{-12}$&$1.0\time10^{-14}$\\
$\nu_\mu$ &$1\times 10^{-12}$&$1.0\time10^{-14}$\\
$\nu_\tau$&$1\times 10^{-12}$&$1.0\time10^{-14}$\\
$u$       &$0.003$           &$3.3\time10^{-5}$\\
$d$       &$0.006$           &$6.6\time10^{-5}$\\
$c$       &$1.3$             &$1.4\time10^{-2}$\\
$s$       &$0.1$             &$1.1\time10^{-2}$\\
$t$       &$175$             &$1.9$\\
$b$       &$4.3$             &$4.7\time10^{-2}$\\
Higgs     &$\approx 150$     &$1.64$\\[1ex] 
\hline 
\end{tabular}
\caption{\textit{Standard Model Particles and Masses}}
\label{table:nonlin} 
\end{table}

\end{document}